\documentclass[prl,twocolumn,showpacs,lengthcheck]{revtex4}

\usepackage{amsmath}
\usepackage{amssymb}
\usepackage{graphicx}

\newcommand{\tr}{\operatorname{tr}}

\begin{document}

\bibliographystyle{apsrev}

\title{Entropic Bounds for the Quantum Marginal Problem}

\author{Tobias J. \surname{Osborne}}
\email[]{Tobias.Osborne@rhul.ac.uk} \affiliation{Department of
Mathematics, Royal Holloway, University of London, Egham, Surrey,
TW20 0EX, UK}

\begin{abstract}
The quantum marginal problem asks, given a set of reduced quantum
states of a multipartite system, whether there exists a joint
quantum state consistent with these reduced states. The quantum
marginal problem is known to be hard to solve in general as it is a
variant of the \emph{$N$-representability problem}. We provide
entropic bounds on the number of orthogonal solutions to the quantum
marginal problem.
\end{abstract}


\maketitle

The quantum marginal problem (QMP) asks when and what joint quantum
states of a composite system are compatible with a given set of
reduced states. This question has its genesis in the \emph{marginal
problem} of classical probability theory which is concerned with the
existence of a probability density function with given projections
onto a set of coordinate subspaces (see \cite{klyachko:2002a} and
references therein). Since the foundational work of
\cite{linden:2002a, linden:2002b}, which revealed that almost all
tripartite quantum states are uniquely determined by their two-party
reduced states, there has been a great deal of progress on
variations and generalisations of the QMP \cite{higuchi:2003a,
higuchi:2003b, bravyi:2004a, han:2005a, klyachko:2004a, franz:2002a,
christandl:2006a, daftuar:2004a, hall:2007a, liu:2007a,
eisert:2007a}. At the current time only the most fundamental version
of the QMP, which asks what constraints prevent the existence of a
joint state $\rho_{A_1A_2\cdots A_n}$ given only the single-particle
reduced states $\rho_{A_j}$, is well understood. This problem was
completely solved by Klyachko in the finite-dimensional setting
\cite{klyachko:2004a}, and by Eisert et.\ al.\ \cite{eisert:2007a}
in the gaussian setting.

A general solution to the QMP would have profound and revolutionary
consequences for physics as it would provide a solution to the
\emph{$N$-representability problem} of quantum chemistry
\cite{coleman:2000a, liu:2007a}, and hence allow us to easily
calculate, eg., the binding energies of complex molecules. It turns
out that this is too much to hope for as a general solution cannot
exist: the $N$-representability problem is now understood to be too
hard to solve, even on a quantum computer \cite{liu:2007a}. Thus, in
order to gain quantitative progress on the QMP we must take recourse
either to approximate or heuristic methods. We take the first
approach here: we derive bounds on the \emph{number} of orthogonal
solutions to the general QMP. We illustrate these bounds in the
tripartite case where we are looking for joint quantum states of a
tripartite system $ABC$ given two reduced states $\rho_{AB}$ and
$\rho_{BC}$ (even the classical marginal problem is unsolved in this
case \cite{klyachko:2002a}).

In this Letter we study a generalisation of the QMP, namely, how
many \emph{orthogonal} solutions are there to the QMP? This problem
is of direct relevance not only to quantum chemistry, but also to
condensed matter physics and quantum complexity theory because
solutions to the QMP arise as \emph{ground states} of locally
interacting hamiltonians, and the number of orthogonal pure-state
solutions is equal to the ground-state degeneracy of the
hamiltonian.

We first focus on the case where we are given two reduced states
$\rho_{AB}$ and $\rho_{BC}$ of a tripartite quantum system $ABC$,
with local dimensions $d_A$, $d_B$, and $d_C$, and we wish to
determine how many orthogonal \emph{pure states} $|\psi_j\rangle$,
$j = 1, 2, \ldots, m$ of $ABC$ are consistent with $\rho_{AB}$ and
$\rho_{BC}$, meaning that $\tr_{C}(|\psi_j\rangle\langle \psi_j|) =
\rho_{AB}$ and $\tr_{A}(|\psi_j\rangle\langle \psi_j|) = \rho_{BC}$,
for all $j$. To approach this problem we suppose that $m$ such
orthogonal states exist and construct the following mixed quantum
state of $ABC$:
\begin{equation}
\rho_{ABC} = \sum_{j=1}^m \frac{1}{m}|\psi_j\rangle\langle\psi_j|.
\end{equation}
Notice that $\rho_{AB} = \tr_{C}(\rho_{ABC})$ and $\rho_{BC} =
\tr_{A}(\rho_{ABC})$.

Our next step is to take our putative state $\rho_{ABC}$ and compute
its von Neumann entropy, $S_{ABC} \equiv
-\tr(\rho_{ABC}\log_2(\rho_{ABC})) = \log_2(m)$, which follows
because the eigenvalues of $\rho_{ABC}$ are $1/m$.

To complete our derivation we then apply the \emph{strong
subadditivity inequality} for the von Neumann entropy
\cite{nielsen:2000a} to $ABC$, which reads $\log_2(m) = S_{ABC} \le
S_{AB} + S_{BC} - S_B$. Thus we obtain the bound
\begin{equation}\label{eq:mbound}
m \le 2^{S_{AB} + S_{BC} - S_B}.
\end{equation}
Note that $S_{AB}$, $S_{BC}$, and $S_B$ are easy to calculate in
terms of the initial data as $\rho_B$ is obtained from $\rho_{AB}$
via partial trace.

By applying the previous argument inductively step we can provide a
general entropic bound on the number $m$ of pure-state solutions to
more general instances of the QMP: suppose we are given the reduced
states $\rho_{A_jA_{j+1}}$, $j = 1, 2, \ldots, n-1$ of an
$n$-partite system (the subsystems $A_j$ need not have the same
dimension), then we can bound the number $m$ of pure-state solutions
to the QMP as follows:
\begin{equation}
\begin{split}
S_{A_1A_2\cdots A_n} &\le S_{A_1A_2\cdots A_{n-1}}+
S_{A_{n-1}A_n} - S_{A_{n-1}} \\
&\quad\quad\vdots\\
&\le \sum_{j=1}^{n-1} S_{A_jA_{j+1}} - \sum_{j=2}^{n-1} S_{A_j},
\end{split}
\end{equation}
where we've repeatedly the strong subadditivity inequality to
$S_{A_1A_2\cdots A_{k}}$ in each step. Thus, using the same
reasoning as above, we have that
\begin{equation}
m \le
\prod_{j=1}^{n-1}2^{S_{A_jA_{j+1}}}\prod_{j=2}^{n-1}2^{-S_{A_j}}.
\end{equation}
It is clear how to extend this argument to more general settings.

The problem of determining the constraints on the existence of a
joint state $\rho_{A_1A_2\cdots A_n}$ given only the single-particle
reduced states $\rho_{A_j}$ is completely solved. However, the
number of such constraints is enormous when the dimensions of $A_j$
become large \cite{klyachko:2004a}. For example, even for the case
of two three-level systems there are 387 inequalities to be checked.
For the problem considered here, even if these inequalities are
satisfied, one still needs to work out how many orthogonal
pure-state solutions there are. Thus it is still desirable to
develop bounds on the number of solutions to the QMP for this
simpler case. We can do this using the \emph{subadditivity} of the
von Neumann entropy as follows. Consider the case where we are given
$\rho_A$ and $\rho_B$: then, as before, the number $m$ of orthogonal
solutions to the QMP is bounded by
\begin{equation}
\log_2(m) = S_{AB} \le S_A+S_B
\end{equation}
so that  $m \le 2^{S_A}2^{S_B}$, and, following the inductive
argument presented above, we have that
\begin{equation}
m \le \prod_{j=1}^n 2^{S_{A_j}}.
\end{equation}

Up to this point we have focussed on the special case where we are
only looking for pure-state solutions to the QMP. However, it could
be that there are no pure-state solutions to the QMP, yet the QMP is
still solvable with a mixed state. To deal with this we extend our
argument to the mixed-state case by using another idea, namely,
\emph{purification} \cite{nielsen:2000a}: any mixed state $\rho$ of
a quantum system $A$ can be realised as the reduced state of a pure
state $|\rho\rangle_{AA'}$ of a ``doubled'' system $AA'$ where $A'$
is a copy of $A'$. To see this write $\rho = \sum_{j=1}^d
p_j|u_j\rangle\langle u_j|$ for the spectral decomposition of
$\rho$. Then
\begin{equation}\label{eq:purific}
|\rho\rangle = \sum_{j=1}^d
\sqrt{p_j}|u_j\rangle_A(|u_j\rangle_{A'}^*)
\end{equation}
is a purification. If we have purifications $|\rho\rangle
=\sum_{j=1}^d \sqrt{p_j}|u_j\rangle_A(|u_j\rangle_{A'}^*)$ and
$|\sigma\rangle=\sum_{j=1}^d
\sqrt{q_j}|v_j\rangle_A(|v_j\rangle_{A'}^*)$ of two mixed states
$\rho$ and $\sigma$ then
\begin{equation}\label{eq:fidoverlap}
\begin{split}
\langle \rho|\sigma\rangle &= \sum_{j,k=1}^n \sqrt{p_jq_k} \langle
u_j|v_k\rangle(\langle u_j|v_k\rangle)^* \\
&= \tr(\sqrt{\sigma}\sqrt{\rho}) = G(\rho, \sigma),
\end{split}
\end{equation}
where $G(\rho, \sigma)$ is a quantity related to the \emph{fidelity}
$F(\rho,\sigma) = \tr(\sqrt{\sqrt{\rho}\sigma\sqrt{\rho}})$ between
$\rho$ and $\sigma$.

We apply this observation in the following way. Suppose that
$\rho_{ABC}$ is a mixed state solution to the QMP where we are given
$\rho_{AB}$ and $\rho_{BC}$. Then let $|\rho_{ABCA'B'C'}\rangle$ be
the purification Eq.~(\ref{eq:purific}) of $\rho_{ABC}$ onto the
doubled system $ABCA'B'C'$. Suppose that there are $m$ such
mixed-state solutions with corresponding purifications
$|\psi_j\rangle$, and suppose that these purifications are
orthogonal. (This orthogonality implies that the supports of
$\rho^{(j)}_{ABC}$ and $\rho^{(k)}_{ABC}$, $j\not=k$, are orthogonal
so that the corresponding solutions $\rho^{(j)}_{ABC}$ to the QMP
have zero pairwise fidelity: $F(\rho^{(j)}_{ABC}, \rho^{(k)}_{ABC})
= \delta_{jk}$.) We then apply our main argument to $|\psi_j\rangle$
to find
\begin{equation}
\begin{split}
\log_2(m) &= S_{ABCA'B'C'} \le 2S_{ABC} \\
&= 2S_{AB} + 2S_{BC} - 2S_{B},
\end{split}
\end{equation}
where we've applied the subadditivity of the von Neumann entropy to
$|\psi_j\rangle$ across the bipartition $ABC:A'B'C'$ and exploited
the fact that $S_{ABC} = S_{A'B'C'}$. So we learn that
\begin{equation}\label{eq:msbound}
m \le 2^{2S_{AB}}2^{2S_{BC}}2^{- 2S_{B}}.
\end{equation}
The extension to the general QMP is clear.

Note that the arguments presented here don't depend on the dimension
of the composite system and hence generalise, in the appropriate
limits, to infinite-dimensional systems.

Physically, the bound Eq.~(\ref{eq:mbound}) says that, in order for
there to be $m$ solutions to the tripartite QMP the entropy, and
hence, our ignorance, of the reduced states of $AB$ and $BC$ needs
to be large enough to suppress the entropy of the interface system
$B$; there needs to be enough room to move at the interface
subsystem to marry up the two reductions into a larger consistent
state. This is intuitively reasonable and provides a physical
interpretation for the following simple \emph{entanglement monogamy}
result \cite{coffman:2000a}. Suppose we are given $\rho_{AB} =
|\Psi^-\rangle_{AB}\langle \Psi^-|$ and $\rho_{BC} =
|\Psi^-\rangle_{BC}\langle \Psi^-|$, where $|\Psi^-\rangle =
\frac{1}{\sqrt{2}}(|01\rangle -|10\rangle)$ is the
spin-$\frac{1}{2}$ singlet, then there are no pure joint states
$\rho_{ABC}$ consistent with these reduced density operators
\cite{coffman:2000a}: the bound Eq.~(\ref{eq:mbound}) reads $m \le
\frac12$, so that $m=0$. Our bound Eq.~(\ref{eq:msbound}) actually
shows much more as we learn that this situation is robust against
perturbations; there must be an ball of states around
$|\Psi^-\rangle_{AB}$ and $|\Psi^-\rangle_{BC}$ where there are no
joint states consistent with both $\rho_{AB}$ and $\rho_{BC}$.

The bounds Eq.~(\ref{eq:mbound}) and Eq.~(\ref{eq:msbound}) are
likely to be tight in the low-entropy regime, where the entropy of
the interface system $B$ is larger than that of $AB$ or $BC$. In
this case the bounds provide an easy way to prohibit the existence
of a solution to the QMP. The bound Eq.~(\ref{eq:mbound}) is also
likely to perform well in the high entropy regime as it reproduces
the exact result in the completely mixed case $\rho_{AB} =
\mathbb{I}/d_Ad_B$ and $\rho_{BC} = \mathbb{I}/d_Bd_C$, namely, $m =
2^{d_A+d_B+d_B}$.

In this Letter we have developed an entropic upper bound on the
\emph{number} of orthogonal solutions to the quantum marginal
problem. This bound also provides nontrivial constraints on the
existence of solutions to the general QMP. Using our bound it
possible that new bounds for the ground-state energy and degeneracy
(and hence, the thermodynamic pressure) of local hamiltonians may be
developed. Additionally, it is likely that our bound will be useful
in quantum complexity theory where it should be able to provide
nontrivial bounds on quantum counting problems.

\acknowledgements

I would like to thank Henry Haselgrove and Michael Nielsen for
helpful conversations which directly inspired this work. This work
was supported, in part, by the University of London central research
fund.

\end{document}